\documentclass{article} 
\usepackage[
colorlinks=true,
linkcolor=blue,
filecolor=blue,
urlcolor=blue,
citecolor=blue,
breaklinks=true,
pdfstartview=Fit,
]{hyperref}
\usepackage{geometry}
\usepackage{graphicx}
\usepackage{cite}
\usepackage{amsmath}
\usepackage{pifont}
\usepackage{indentfirst}
\usepackage{float}
\usepackage{xcolor}
\usepackage{authblk}
\usepackage{abstract}
\usepackage{caption}

\title{Spatio-spectral light modulator of XUV high harmonics}
	\author[1]{Qi Zeng}
	\author[1]{Yimin Deng}
	\author[1]{Xinyue Yang}
	\author[1]{Wei Cao \thanks{weicao@hust.edu.cn}}
	\author[1,2]{Peixiang Lu}
	\affil[1]{School of Physics and Wuhan National Laboratory for Optoelectronics, Huazhong University of Science and Technology, Wuhan 430074, China}
	\affil[2]{Optics Valley Laboratory, Wuhan 430074, China}

\date{}

\begin{document}
\maketitle

\vspace{-5em}
	\begin{abstract}
High-order harmonic generation (HHG), characterized by its highly nonlinear nature, often exhibits a complex spatio-temporal profile that poses challenges for practical applications. In this study, we unveil a method for manipulating the spatio-spectral distribution of HHG by guiding the recollision electron trajectory in the spatio-temporal domain using a control field. The resulting far-field high harmonic (HH) radiation inherits the intricate spatio-temporal characteristics of the control field, showcasing diverse features including spatial tilting, spectral shifting, and emission angle deflection. Using the relative delay between the control field and the driving pulse as the primary control parameter, we achieve precise tailoring of the high harmonics in the spatio-spectral domain. This controllability in HH benefits ultrafast metrology and imaging applications in the extreme ultraviolet (XUV) regime.
	\end{abstract}

\section{Introduction}

\par High-order harmonic generation is a non-perturbative nonlinear process, in which the fundamental driver in the near-infrared (NIR)/visible is frequency up-converted to the XUV/soft X-ray regime. It serves as a high quality tabletop coherent XUV/soft X-ray source that holds promise for probing and even imaging ultrafast dynamics. Due to the highly nonlinear nature of HHG, the plasma-induced field distortion and intensity dependent dipole phase may lead to complex spatio-temporal structure in the HH radiation \cite{RN135,RN136}. Although HH has been successfully utilized to probe electron dynamics in atoms, molecules, and solids \cite{RN127,RN128,RN129,RN130,RN131,RN132,RN133,RN134}, the dynamics retrieved so far are typically space averaged. Advancing the ultrafast measurement to incorporate the spatial resolution necessitates the spatio-temporal control of HH radiation. Directly engineering optical components in the XUV/soft X-ray regime remains challenging due to the materials' strong absorption, this promotes extensive investigations of XUV beam manipulation based on gas phase medium. For instance, the significant refractive index change across a transition resonance allows controlling the XUV pulse wavefront using a gas jet with a density gradient as linear XUV optics \cite{RN140}. In addition, the XUV wavefront can also be manipulated via nonlinear processes. L. Drescher et al. carefully designed the phase matching of a four-wave mixing process and achieved spectral focusing of broadband XUV radiation in a krypton gas jet \cite{RN141}. Bengtsson et al. utilized an infrared-induced Stark effect to control the spatial phase of the free induction decay after XUV excitation, enabling steering of the deflection angle of an XUV light \cite{RN143}. In this ex-situ category of wavefront manipulation, the XUV source is generated and controlled in two separate gas jets in an extended beam path. Alternatively, an in-situ manner for wavefront manipulation, in which the generation and control of the XUV source is achieved in a single gas jet, is also demonstrated under a rather compact geometry. By introducing a weak field to perturb the HHG process, the radiation dipole will accumulate an additional phase and either the emission direction or the frequency of the high harmonics can be harnessed \cite{RN144,RN145,RN156}. However, previous studies have primarily focused on manipulating a single aspect of HH radiation, either spatial or spectral phase. Simultaneous control over spatial and spectral phases at high precision remains unexplored. 

\par Here, we propose and experimentally demonstrate an all-optical method to control the spatio-spectral properties of HH. A weak NIR control light is combined with a strong laser field to produce high harmonics. Experimental measurements and quantum simulations confirm that the delay/spatial dependent 2D map of the HH spectral shift resembles the spatio-temporal electric field of the control light. This has provided a rather precise and handy way for spatio-spectral tailoring of high harmonics. By varying the delay between the control light and the driving field, we observe that the HH location can be precisely positioned in the spatio-spectral domain with mrad-meV precision. Theoretical analysis suggests that this phenomenon can be interpreted by few-slit interference in the space-time domain.

	\section{Principle and Method}

	\begin{figure*}[ht!]
	\includegraphics[width=1\linewidth]{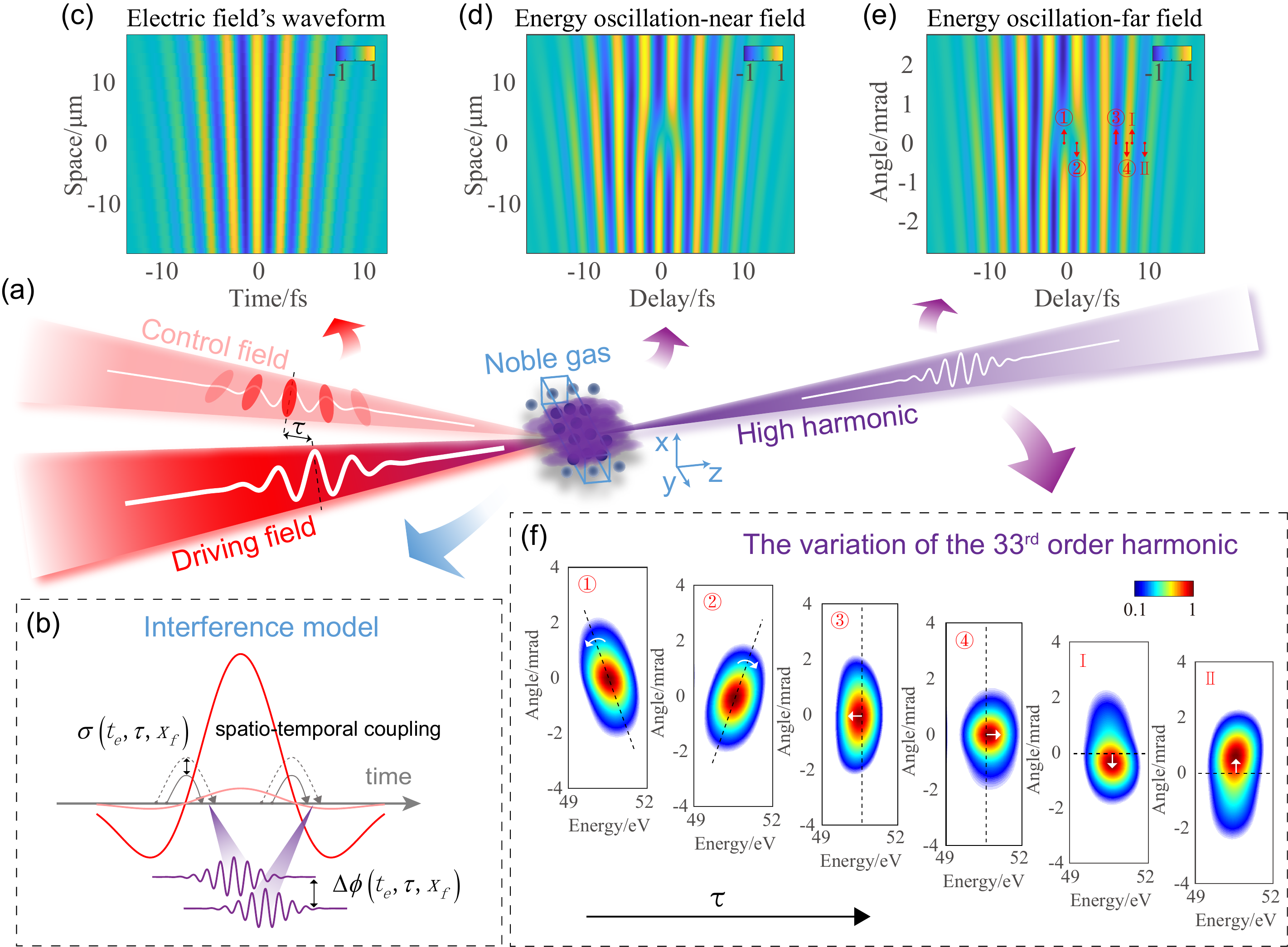}
	\caption{\label{Figure 1} \footnotesize (a) The schematic diagram of the spatio-spectral light modulator of HH. The driving field generates HH after interacting with the gas, and the spatio-temporally coupled control field modulates the HHG process, then the spatio-spectral profile of HH can be modulated by changing the delay of the control field. (b) The principle of spatio-spectral light modulator of HH. The attosecond pulse train generated by the strong driving field is equivalent to a temporal multi-slit. The relatively weak control field introduces an additional spatio-temporally coupled dipole phase to each attosecond slit, which leads to modulation of a high harmonic spectrum in spatio-spectral domain. (c) The electric field of the control field. (d) and (e) are the two dimensional spectral shifts of the $33^{rd}$ order harmonic in the near and far fields, respectively. (f) The spatio-spectral beam profiles of the $33^{rd}$ order HH at different delays indicated in (e), the black dashed lines for \ding{174}, \ding{175}, \uppercase\expandafter{\romannumeral1}, and \uppercase\expandafter{\romannumeral2} specify the centroid of the harmonic without the control field.}
	\end{figure*}

\par Fig.\ref{Figure 1} shows the principle of spatio-spectral tailoring of HHG. A strong NIR femtosecond laser is used to drive high harmonics. When a time-delayed weak control field is introduced to perturb HHG, an extra phase will be imposed on the dipole moment. Previous studies \cite{RN154} considered the introduced control field to be spatially averaged and only the temporal evolution was considered. However, when the control field is inhomogeneous or even spatio-temporally coupled, the spatial part cannot be neglected. Ionization time and radiation time of HH exist in pairs under the saddle point model \cite{RN155} and the induced phase shift can be expressed by (all the formula derivation is presented in supplementary section 1):
\begin{equation}\label{equ1}
	\begin{aligned}
\sigma^j \left( {{t_e},\tau ,{x_f}} \right) &= {S_1}\left( {{t_e},\tau ,{x_f}} \right) - {S_0}\left( {{t_e},{x_f}} \right)\\
&\approx - \frac{9}{{128}}{e_d}\omega_d{t_e^4}E_c\left( {\tau + {\zeta _\varepsilon} ,x_f} \right),
	\end{aligned}
\end{equation}
where the superscript $j$ denotes the corresponding quantum trajectory, $S_1$($S_0$) is the classic action with (without) the control field, $t_e$ is the emission time of HH, $\tau$ is the delay between the driving pulse and the control field, $x_f$ is the space coordinate at the focus, $\omega_d$ ($e_d$) is the frequency (amplitude) of the driving field, $E_c$ is the electric field of the control light, and $\zeta _\varepsilon$ represents a constant associated with harmonic order. When the driving field interacts with the atoms, HHs are radiated once every half optical cycle of the driving field, thus multi-slit interference occurs in the time domain, see Fig.\ref{Figure 1}(b). For simplicity, we only consider two slits in one optical cycle. Considering that the electric fields of the driver for neighboring time slits have opposite signs, the perturbing phases satisfy: ${\sigma _1^j}\left( {{t_e},\tau ,{x_f}} \right) =  - {\sigma _2^j}\left( {{t_e},\tau ,{x_f}} \right)$. When the HHs radiated from two adjacent attosecond slits are in constructive interference, the total phase difference $\Delta \phi \left( {\omega ,\tau ,{x_f}} \right)$ is:
\begin{equation}\label{equ2}
	\begin{aligned}
{\sigma _2^j}\left( {{t_e},\tau ,{x_f}} \right) - {\sigma _1^j}\left( {{t_e},\tau ,{x_f}} \right) + \omega {T_0}/2 + \pi  + \gamma  = 2m\pi,
	\end{aligned}
\end{equation}
where $T_0$ is optical cycle of the driving field, $\omega$ is the frequency of HH, $\gamma$ is the non-adiabatic effect constant \cite{RN157}, and $m$ is an integer. The extra phase terms $\sigma _1^j$ and $\sigma _2^j$ in Eqn.\ref{equ2} alter the constructive interference conditions and thus lead to a spectral shift of a specific harmonic order: 
	\begin{equation}\label{equ_a}
	\delta\omega(\tau,{x_f}) \propto {\alpha(x_f) {E_c}\left( {\tau  + \Delta(x_f) ,{x_f}} \right) + {E_c}\left( {\tau  ,{x_f}} \right)},
	\end{equation}
where $\alpha(x_f)$ is the intensity ratio of the driving field between two neighboring slits, and $\Delta(x_f)$ is the time difference between the two slits. This simple interference model shows that the spectral shift of the HH in the ($\tau$, $x_f$) domain resembles the electric field of two identical control lights separated by $\Delta(x_f)$. 

\par To validate our method, we carried out numerical simulation under the strong field approximation framework \cite{RN160}, and the thin gas medium \cite{RN161} assumption was also utilized.  As an example, we used an Electromagnetic field with wavefront rotation (WFR) as the control pulse, in fact, the WFR effect is equivalent to the spatial chirp effect \cite{RN148}: ${{\rm{E}}_c}\left( {t ,{x_f}} \right) = {e_c}\cos \left[ {\left( {{\omega _c} + \eta {x_f}} \right)t } \right]$, where $\eta  = 4\frac{{{w_i}\xi }}{{{w_f}{T_f}{T_i}}}$, $e_c (\omega _c)$ is the amplitude (frequency) of the control field, $w_i$ ($T_i$) is the spot size (duration) of the control field before focusing, $w_f$ ($T_f$) is the spot size (duration) of the control field at the focus, $\xi=\rho\xi_0$, $\rho$ is a constant, and $\xi_0$ is a spatio-temporal parameter. The spatio-temporal waveform of the control field is shown in Fig.\ref{Figure 1}(c). To mimic the conditions of short quantum trajectory phase-matching, we applied super-Gaussian time windows on the dipole moment to select short quantum trajectories \cite{RN162} (see supplementary section 2 for details). Fig.\ref{Figure 1}(d) shows the two dimensional spectral oscillation of the $33^{rd}$ order HH in the near field. Its pattern consists of a translation, scaling, and superposition of Fig.\ref{Figure 1}(c). It can be seen that this 2D image is directly linked to the electric field of the control light as predicted by Eqn.\ref{equ_a}. The far field distribution of HHs can be calculated using the Fraunhofer diffraction formula. Notably, our simulation findings show that the delay and spatial dependent spectral shift of a specific harmonic order demonstrates a rather similar pattern for both near and far fields, see Fig.\ref{Figure 1}(d) and (e). Thus the far-field modulation of the HHs can be written as: 
	{\begin{equation}\label{equ2_2}
	\delta\omega(\tau,\theta) \propto {\alpha(\theta) {E_c}\left( {\tau  + \Delta(\theta) ,{\theta}} \right) + {E_c}\left( {\tau  ,{\theta}} \right)}.
	\end{equation}}

\par When the control field exhibits spatio-temporal coupling, it modulates the HHG process and induces an extra phase $\sigma$ carrying the spatio-temporal coupling characteristics in the attosecond pulses radiated at different moments. As a result, the attosecond pulse trains interfere and propagate to the far field through diffraction, transferring the spatio-temporal properties of the control field to the HHs in the far field. Using the equation presented above, we establish a quantitative relationship between HH modulation in the spatio-spectral domain and the control field. This allows for the design and manipulation of pulses in a more readily accessible optical regime. By carefully designing the control field, we can achieve tailored modulation of the XUV light field to meet various application needs.

\par The simulated spatio-spectral profiles of the $33^{rd}$ order HH at different delays are shown in Fig.\ref{Figure 1}(f). These specific delays are labeled as \ding{172}-\ding{175} and \uppercase\expandafter{\romannumeral1}-\uppercase\expandafter{\romannumeral2} in Fig.\ref{Figure 1}(e). As the delay varies, the harmonic shows diverse features including spectral shifting (\ding{174},\ding{175}), emission angle deflection (\uppercase\expandafter{\romannumeral1},\uppercase\expandafter{\romannumeral2}), and even orientation tilt (\ding{172},\ding{173}). The harmonic modulation mechanism is mainly related to the pattern of Fig.\ref{Figure 1}(e). It presents that different emission angles of the HH have different spectral shifts. The color bar in Fig.\ref{Figure 1}(e) indicates the magnitude of the spectral modulations, with a positive value representing a blue shift and a negative value indicating a red shift. Specifically, for delays \ding{172}-\ding{173}, the spectral shift of the HH exhibits a sign change from the positive emission angle to the negative emission angle, meaning that the harmonic bears a spatial chirp character. This spatial chirp corresponds to a rotation of the harmonic beam profile in the spatio-spectral domain. As the delay of the control field varies, the spatial chirp quantity changes, leading to different rotation angles of the harmonics. For delays \ding{174}-\ding{175}, the HH experiences overall blue and red shifts, respectively, but there are no significant spatial modulations because the spectral and spatial shifts are not synchronized; a phase difference prevents their alignment, as will be further discussed in the Discussion section. Therefore, the spatio-spectral structure of the HHs is encoded in Fig.\ref{Figure 1}(f) through Eqn.\ref{equ2_2}. For delays \uppercase\expandafter{\romannumeral1}-\uppercase\expandafter{\romannumeral2}, the spectral shift varies much more slowly with space coordinates as compared to \ding{172}-\ding{173}, resulting in a lack of noticeable rotational effects of the HH spatio-spectral profile. However, profound emission angle deflections are observed, primarily due to the perturbing phase introduced by the control field, and will be addressed later in the Discussion section. Therefore, the control light can act as a frequency shifter, a deflecting optics, or a spatial-chirping element depending on the delay between the driving pulse and the control field, which can be regarded as an XUV spatio-spectral light modulator. It should be noted that although the $33^{rd}$ order harmonic is selected for demonstration, other harmonics show rather similar patterns and therefore this method is independent of harmonic order, and the operational principle of this method roots on Eqn.\ref{equ2_2}. Although a first order spatio-temporal coupling control light is used for demonstration, other forms of an electric field such as pulse front tilt (PFT) pulse, and noncollinear plane wave can also be utilized.

\section{Experimental result}

	\begin{figure*}[ht!]
	\centering
	\includegraphics[width=1\linewidth]{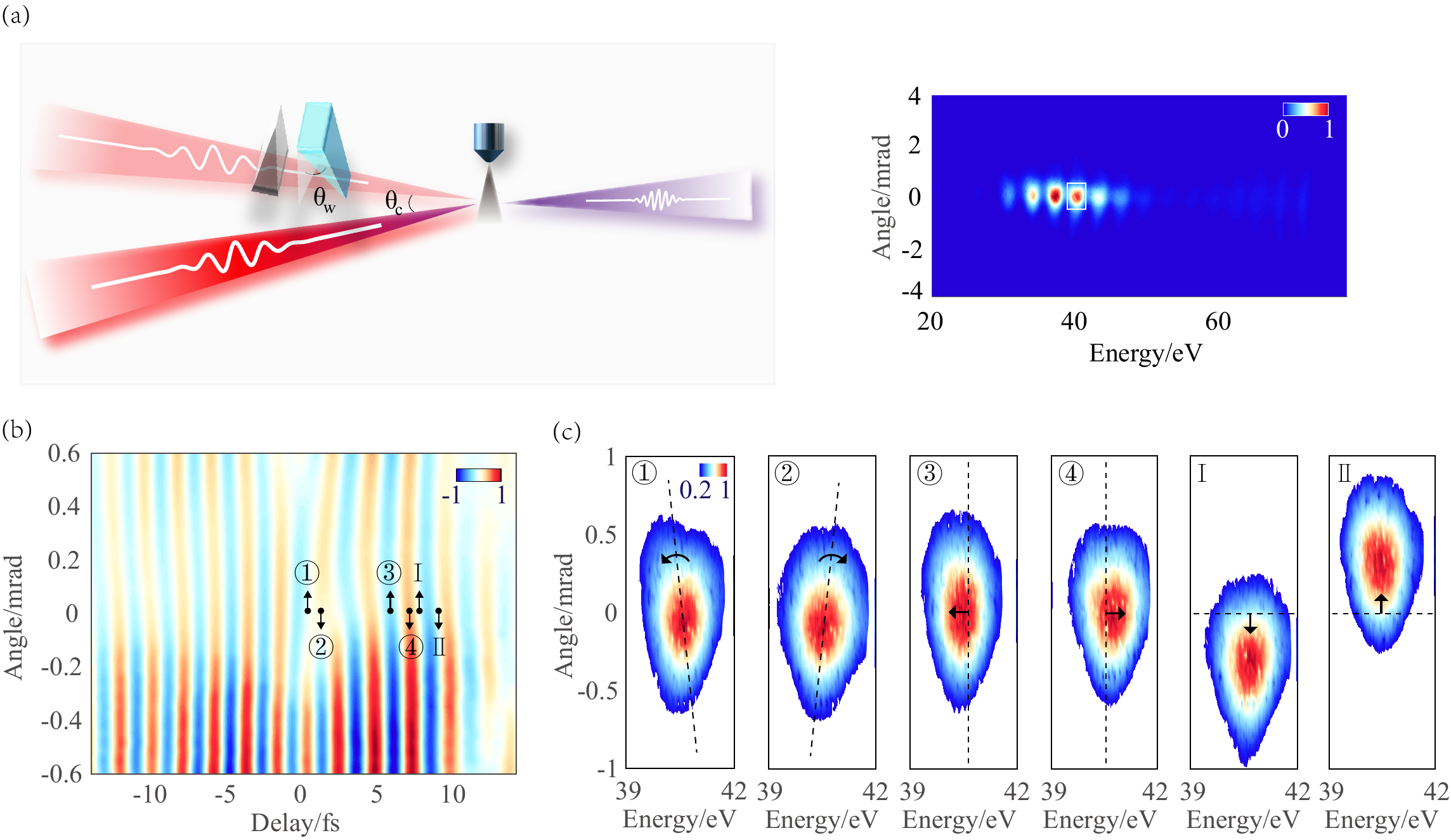}
	\caption{\label{Figure 2} \footnotesize (a) The simplified schematic diagram of the experiment, where the PFT effect is introduced by manipulating the control field with a set of misaligned wedge pairs. The plot on the right-hand side is the high harmonic spectrum captured using an XUV spectrometer, with the white box highlighting the $25^{th}$ order HH. (b) Measured two dimensional spectral shift of the  $25^{th}$ order HH. (c) Spatio-spectral beam profiles of the $25^{th}$ order HH at different delays indicated in (b).}
	\end{figure*}

\par Our experimental setup is based on the Mach-Zehnder interferometer (see supplementary section 3). The driving field (few-cycle, the wavelength is 780 nm, intensity is estimated to $3.5 \times {10^{14}}$ $\mathrm{W/c{m^2}}$) interacts with argon gas injected from a pulsed nozzle with an aperture of 220 ${\mu}$m and a back-pressure of 20 bar to generate high harmonics. The short trajectories can be selected by properly positioning the gas jet with respect to the driving laser focus \cite{RN103}. A replica of the driving pulse is reflected by a beam splitter to serve as the control field whose incidence angle can be adjusted by controlling the angle of the beam splitter. Its intensity is estimated to be $9 \times {10^{11}}$ $\mathrm{W/c{m^2}}$. The PFT effect in the control field can be introduced by inserting a pair of misaligned wedges (top angle is $8^{\circ}$, misaligned angle $\theta_w$ is $30^{\circ}$) in the beam path, see the left side of Fig.\ref{Figure 2}(a). After focusing by a concave mirror, the pulse will carry WFR at the focus \cite{RN148}. By adjusting the nanometer-precision delay stage (P-622.1CD, Physik Instrumente), we can change the delay and modulate the high-order harmonic generation process with attosecond precision. An optical spectrometer (FLAME-T-VIS-NIR, Ocean Optics) is used to measure the spatial spectrum of the control field at the laser-gas interaction region, and the results (see supplementary section 4) show a clear spatial chirp effect, which is consistent with previous results \cite{RN150}. The spectrum of the HHs is detected by an extreme ultraviolet spectrometer, see the right side of Fig.\ref{Figure 2}(a). We analyze the $25^{th}$ order HH and present its spectral shift as a function of delay and emission angle in Fig.\ref{Figure 2}(b). Fig.\ref{Figure 2}(b) and Fig.\ref{Figure 1}(e) reveal a similar structure, particularly a fork-like pattern that appears near zero delay. To further investigate, we select six specific delay moments and analyze the corresponding HH features. Our results show that near zero delay, the HHs exhibit rotational behavior, see Fig.\ref{Figure 2}(c) \ding{172}-\ding{173}, while as the delay increases, the HHs show spectral and spatial shifts, see Fig.\ref{Figure 2}(c) \ding{174}-\ding{175} and \uppercase\expandafter{\romannumeral1}-\uppercase\expandafter{\romannumeral2}. These findings are in good agreement with our simulation results in Fig.\ref{Figure 1} confirming the validity of our approach.

\section{Discussion}

	\begin{figure*}[ht!]
	\centering
	\includegraphics[width=1\linewidth]{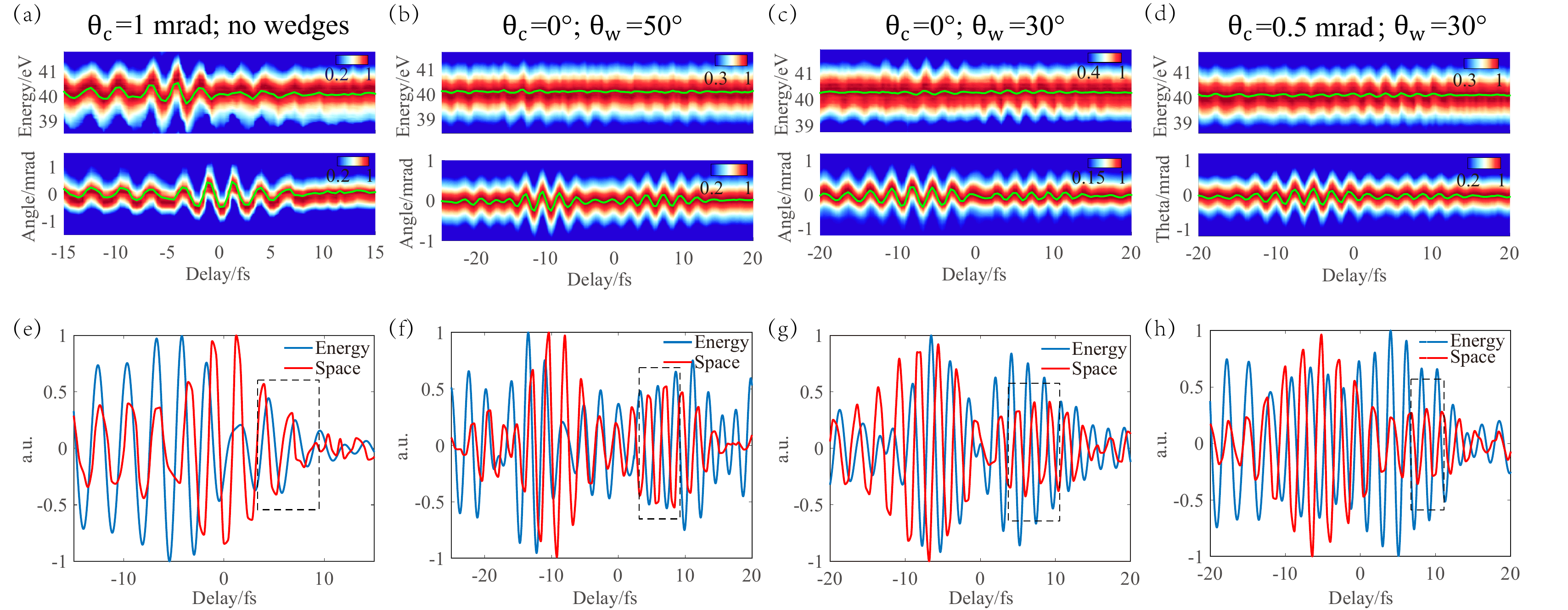}
	\caption{\label{Figure 3} \footnotesize  (a)-(d) The spatially integrated spectrum (upper row) and spectrally integrated spatial distribution (lower row) of the $25^{th}$ order HH under different control field conditions in the experiment. The green solid lines show the center of mass of the modulation. $\theta_c$ represents the angle between the control field and the driving pulse, and $\theta_w$ represents the misaligned angle of the wedge pairs. (e)-(h) The normalized center of mass curves from (a)-(d). The spectral shift (blue solid lines) of HH oscillates non-synchronously with its spatial shift (red solid lines), as predicted by Eqn.\ref{equ3_1} and Eqn.\ref{equ3_4}.}
	\end{figure*}

\par The above results illustrate how the control field, which carries a WFR effect, is used to modulate high-order harmonic generation in both simulations and experiments. The comparison highlights the rich features of this modulation technique in the spatio-spectral domain, demonstrating that the spatial tilt, spectral shift, and emission angle of the HHs can be precisely controlled through the control field. This all-optical modulation method only requires the design of the spatio-temporal characteristics of the control field, with the HHG distribution in the spatio-spectral domain being precisely adjustable by varying the delay between the control field and the driving pulse. Due to the complex nature of this modulation, we integrate the HH across both the spectral and spatial dimensions to extract its center of mass position in the spatio-spectral domain for quantitative analysis, which reveals that our approach can achieve spatio-spectral modulation accuracy on the order of mrad-meV.
 
\par The delay dependent spectral shift of a specific harmonic order can be derived from Eqn.\ref{equ2_2} after performing spatial integration as: 
\begin{equation}\label{equ3_1}
\delta \omega \left( \tau  \right)  \propto \alpha {E_c}\left( {\tau  + \Delta } \right) + {E_c}\left( \tau  \right).
\end{equation}
This formula remains valid for the first-order spatio-temporal coupling or noncollinear control fields (see supplementary section 5). The delay dependent emission angle of a specific harmonic order is encoded in the dipole phase. To the lowest order approximation, the averaged emission angle of a harmonic can be evaluated as the first order derivative of the spatial phase with respect to spatial coordinate: $\delta \theta \left( \tau  \right) \propto \frac{{\partial {\sigma ^{{j}}}\left( {\tau ,{{{x}}_f}} \right)}}{{\partial {{{x}}_f}}}$. Any spatio-temporal coupling or noncollinear incidence in the control field leads to a nonzero spatial derivative of the HH dipole phase. Therefore, the radiation angle of the HH is modulated by the control field. This modulation explains why the HHs exhibit spatial shift with delay, as illustrated in Fig.\ref{Figure 1}(f)\uppercase\expandafter{\romannumeral1}-\uppercase\expandafter{\romannumeral2} and Fig.\ref{Figure 2}(c)\uppercase\expandafter{\romannumeral1}-\uppercase\expandafter{\romannumeral2}. When the control field is a noncollinear plane wave, the averaged deflection angle: 
\begin{equation}\label{equ3_2}
\delta \theta\left( \tau  \right)  \propto {A_c}\left( \tau  \right), 
\end{equation}
where $A_c$ is the vector potential of the control field. For a collinearly propagating control field with WFR, the equation is 
\begin{equation}\label{equ3_3}
\delta \theta\left( \tau  \right)  \propto \eta \tau {A_c}\left( {\tau  + {\varphi _c}} \right), 
\end{equation}
where $\varphi_c=-0.196\pi$ is the phase shift due to the introduction of spatio-temporal coupling effects, and the value of this constant phase is obtained by a trigonometric curve fitting (the fitting results are presented in supplementary section 6). For the case of a noncollinear propagating control field with WFR, the equation turns to: 
\begin{equation}\label{equ3_4}
\delta \theta\left( \tau  \right)  \propto \kappa {A_c}\left( \tau  \right) + \eta \tau {A_c}\left( {\tau  + {\varphi _c}} \right), 
\end{equation}
where $\kappa$ is proportional to the noncollinear angle $\theta_c$ and specifies the relative contribution of the noncollinear effect to the emission angle of HHs. Eqn.\ref{equ3_1} and \ref{equ3_4} indicate that there exists a phase difference between $\delta \omega$ and $\delta \theta$, the spectral and spatial oscillations as a function of delay are not synchronized, as illustrated in Fig.\ref{Figure 1}(f)\ding{172}-\ding{175} and Fig.\ref{Figure 2}(c)\ding{172}-\ding{175}.

	\begin{figure*}[ht!]
	\centering
	\includegraphics[width=1\linewidth]{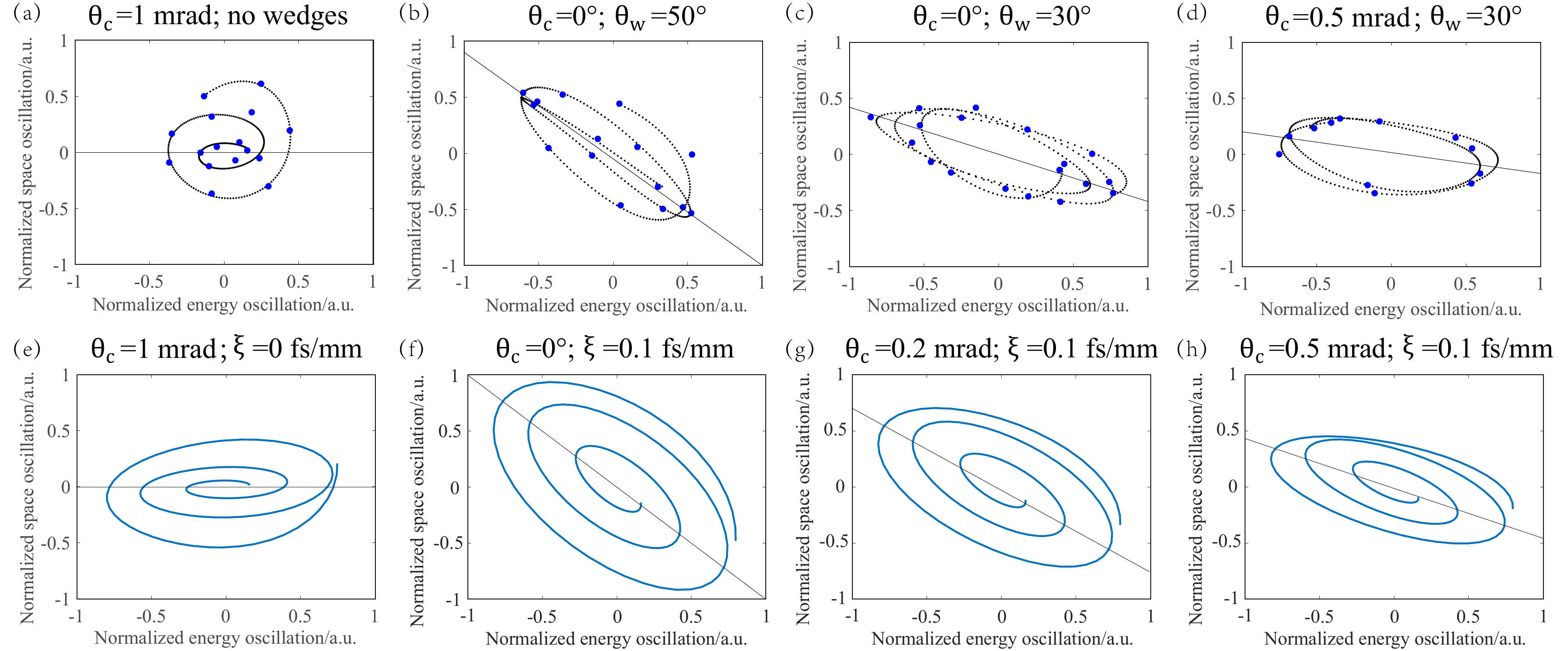}
	\caption{\label{Figure 4} \footnotesize Ellipses are plotted using the extracted center of mass of the spatially integrated spectrum and spectrally integrated spatial distribution as two variables. (a)-(d) display the experimental results under different control field conditions. (e)-(h) show the simulation results under different control field conditions.}
	\end{figure*}

\par We extract the center of mass of the spatially and spectrally integrated spectrum of the $25^{th}$ order HH respectively, and both of them are normalized and plotted as a function of delay for four different control fields, as shown in Fig.\ref{Figure 3}. Under the condition of the WFR effect, from Eqn.\ref{equ3_1}, it can be seen that the delay dependent spectral shift is simply the superposition of two replicas of the control field separated by $\Delta$. While Eqn.\ref{equ3_3} shows that the emission angle of HHs can be expressed as the product of the vector potential of the control field and the delay. This means that the delay dependent emission angle goes across zero as $\tau=0$, forming a dual-pulse pattern centered around zero delay. The above conclusion is in correspondence with the experimental results, see Fig.\ref{Figure 3}(f) and (g), and verifies the correctness of our model. When the noncollinear propagating effect is combined with the WFR effect, the ratio of modulation amplitudes between the positive and negative delay regions is altered in accordance with Eqn.\ref{equ3_4} and is proportional to $(k-\eta)/(k+\eta)$. When the value of WFR $\eta$ is fixed, increasing the noncollinear angle $\theta_c$ (or $k$) will suppress the modulation amplitude for positive delays. This trend of change is consistent with the experiment, see Fig.\ref{Figure 3}(g) and (h). Please note that a perfect collinear propagation of the driving and control field is difficult, we estimate a small noncollinear angle of $ \sim \pm0.2$ mrad in our experiment even under the collinear condition, which makes the amplitudes of the double pulses unequal in Fig.\ref{Figure 3}(f) and (g). The shorter modulation duration observed in the noncollinear experiment in Fig.\ref{Figure 3}(a), compared to the other experimental conditions, is due to the absence of an additional set of wedge pairs, which provides a shorter control field with less dispersion.

\par The delay dependent spectral shift $\delta \omega$ can be plotted against the delay dependent emission angle $\delta \theta$ for positive delay regions indicated by the black dashed boxes of Fig.\ref{Figure 3}(e)-(h), and the results are shown in Fig.\ref{Figure 4}(a)-(d). From the above equation, for the positive delay, the spectral shift is proportional to $E_c(\tau)$, and the spatial shift is proportional to $A_c(\tau,\varphi _c)$, $\varphi _c$ is 0 for noncollinear configuration, $\varphi _c \ne 0$ for WFR. Therefore, when the condition of the control field is only noncollinear, it presents a regular ellipse, see Fig.\ref{Figure 4}(a). While the WFR effect is present in the control field, due to the additional phase shift $\varphi _c$ introduced, ($\delta \omega$,$\delta \theta$) follows a tilted ellipse (see Fig.\ref{Figure 4}(b)-(d)). As the noncollinear angle $\theta _c$ increases, the tilting angle of the ellipse will decrease.

\par Numerical simulations based on SFA and thin slab model are performed and the parameters are similar to the experimental conditions while ignoring the propagation effect. Indeed, in our experiments, we utilize a pulsed nozzle that injects a very thin layer of gas, with the pressure at the nozzle exit only around a few torr, enabling the neglect of propagation effects. The driving field duration is 8 fs and intensity is $2\times {10^{14}}$ $\mathrm{W/c{m^2}}$; the intensity of the control field is $5\times {10^{11}}$ $\mathrm{W/c{m^2}}$; and the interacting gas is argon. A relatively higher order of harmonic ($33^{rd}$) is chosen for analysis since it can achieve better short trajectory selection by applying a super-Gaussian temporal window on the dipole moment. Using the same method applied to the experimental data, we analyze the simulated HHs spectra under various control fields. The integrated spectral and spatial modulations are presented in section 7 of the supplementary material. Additionally, the extracted delay dependent spectral and spatial shift of $33^{rd}$ order harmonic are plotted in Fig.\ref{Figure 4}(e)-(h), showing good agreement with the experimental results. These results demonstrate that by varying the delay between the control light and the driving field, the HH location can be precisely positioned in the spatio-spectral domain with mrad-meV precision.

\par For the sake of rigor, we also carry out further simulations for a more complex control field. In addition to the first-order spatio-temporal coupling effect, we also study the second-order spatio-temporal coupling effect \cite{RN158,RNa}. Since the spatial chirp has a parabolic shape, the shape of the HH takes on a ''crescent'' shape (see supplementary section 8). Different orders of HH are selected to analyze and we can obtain the same conclusion (see supplementary section 9). When a spatio-temporal coupling control field is introduced in a collinear configuration, the ellipse remains essentially unchanged by varying the value of WFR (see supplementary section 10). When we fix the noncollinear angle but change the value of WFR, the larger the value of WFR, the closer the ellipse tilt angle is to $45^{\circ}$ (see supplementary section 11). These results show that by combining the incident angle with the spatio-temporal coupling field, a variety of modulation modes can be achieved. This all-optical method for precisely modulating spectral and spatial properties of HHs is not only feasible but also straightforward, offering extensive modulation capabilities.

\section{Conclusion}
\par In conclusion, we demonstrate an all-optical XUV spatio-spectral modulator, for precisely and simultaneously controlling the spectral and spatial properties of HHs. This approach extends techniques previously applicable only in the optical regime to the XUV. It relies on using a spatio-temporal coupling control field to modulate the HHG process. By altering the spatio-temporal phase of the dipole radiation of HHs, the HHs undergo spatio-temporal interference and are subsequently diffracted into the far field. Since this modulation is straightforward and based on an interference model, it offers higher precision than existing diffraction-based models. Both the frequency and emission angle of the HHs can be controlled with accuracies in the meV and mrad ranges. This technique enables the generation of HHs with various spatial and spectral features, such as spatial tilting, spectral shifting, and emission angle deflection, in both the near and far fields, providing a versatile tool for future applications. Moreover, the method does not require complex instrumentation; it only necessitates a well-designed NIR pulse, which can be generated using common experimental setups. Furthermore, we can manipulate the incident angle of the control field along with its spatio-temporal character to harness high harmonics with complex spatio-spectral properties. The high spatio-spectral modulation accuracy of this approach broadens the range of HH applications and offers deeper insight into the transfer of spatio-temporal coupling information from the control field to the high-order nonlinear process. This understanding contributes to advancements in attosecond metrology and precise control of HH properties.

\section*{Funding}
\par This research was supported by State Key Development Program for Basic Research of China under Grant No.2024YFE0205800 and No. 2023YFA1406800, the National Natural Science Foundation of China under Grant No. 12274158, the Open Foundation Project of Hubei Key Laboratory of Optical Information and Pattern Recognition of Wuhan Institute of Technology under Grant No. 202304.

\section*{Disclosures}
The authors declare no conflicts of interest.

\section*{Data availability}
Data underlying the results are not publicly available at this time but may be obtained from the authors upon reasonable request.

\section*{Supplemental document}
See Supplement 1 for supporting content.

\end{document}


\maketitle

\vspace{-5em}
	\begin{abstract}
We provide some important information to support the feasibility of our work in this supplementary material. In section 1, the equations in the manuscript are derived. In section 2, the simulation of high harmonic spectrum after selecting the short trajectories is shown. In section 3, we present our experimental setup. In section 4, the spatial spectrum of the control field we measured is shown. In section 5, we compare the simulation result of the spectral shift under different conditions of the control field. In section 6, we present the fitting result of the ellipse. In section 7, we present other results of the simulation. In section 8, the modulation of high harmonic with second-order spatiotemporal coupling effect is provided. In section 9, we show that different orders of high harmonic have the same behavior. In section 10, we present the simulation result for the control field in the collinear configuration and with different values of wave front rotation(WFR). In section 11, we present the simulation result for the control field with the same noncollinear angle but different values of WFR.
	\end{abstract}

\section*{Section 1: Derivation of the equations}
	
\par Ionization time and radiation time of high harmonic exist in pairs under the saddle point model \cite{RN155}. When the control field is sufficiently weak, it does not affect the ionization process but mainly perturbs the free electron’s trajectory, the time-dependent dipole moment \cite{RN154} after considering the spatial part can be approximately written as:
	\begin{equation}	\label{equ1}
	d\left( {{t_e},\tau ,{x_f}} \right) = \sum\nolimits_j {d_0^j} \left( {{t_e},{x_f}} \right){e^{ - i{\sigma ^j}\left( {{t_e},\tau ,{x_f}} \right)}} + c.c
	\end{equation}
Where $d_0$ is the dipole without perturbation, the superscript $j$ denotes the corresponding quantum trajectory, $t_e$ is recollision time, $\tau$ is the delay between driving field and control field, and $x_f$ is the space coordinate at the focus. When we introduce the control field to perturb the high-harmonic generation process, an additional phase will be introduced, which can be written as:
	\begin{equation}	\label{equ2}
	{\sigma ^j} (t_e,\tau ,x_f){\rm{ = }}{{\rm{S}}_1}(t_e,\tau ,x_f) - {{\rm{S}}_0}(t_e,x_f)
	\end{equation}
Where $S_1$ ($S_0$) represents the classic action with (without) the control field, respectively. On the account of: 
	\begin{equation}	\label{equ3}
	{{\rm{S}}_0}(t_e,x_f) = \int_{t_i}^{t_e} {dt'} \left[ {{v(t',x_f)^2}/2{\rm{ + }}{I_p}} \right]
	\end{equation}
Where $v$ is the velocity. Eqn.(\ref{equ2}) can be written as:
	\begin{equation}	\label{equ4}
	\begin{aligned}
{\sigma ^j} (t_e,\tau ,x_f)&{\rm{ = }}\int_{t_i}^{t_e} {\left[ {{{{\left( {{v_{d }} + {v_c}} \right)}^2}} /2 + {I_p}} \right]dt}  - \int_{t_i}^{t_e} {\left( { {v_{d }^2 } /2 + {I_p}} \right)dt} \\
&{\rm \approx} -  \int_{t_i}^{t_e} {\left( {{v_{d}}{A_c}} \right)dt} 
	\end{aligned}
	\end{equation}
Where $A_c$ is the vector potential of the control field, $v_{d} (v_{c})$ is the velocity of the quantum trajectory under the driving (control) field. According to the approximation of research \cite{RN159}, we can write Eqn.(\ref{equ4}) as:
	\begin{equation}	\label{equ5}
{\sigma ^j} (t_e,\tau ,x_f) \approx  - \frac{9}{{128}}{e_d}\omega_d{t_e^4}E_c\left( {\tau + {\zeta _\varepsilon} ,x_f} \right)
	\end{equation}
Where $\omega_d$ and $e_d$ are the frequency and the amplitude of the driving field, $\zeta _\varepsilon$ represents a constant associated with harmonic order, which can be neglected when considering a specific harmonic order. For different forms of control fields, we can get different perturbation phases. When we use the control field with WFR \cite{RN148}:
	\begin{equation}	\label{equ6}
{{\rm{E}}_c}\left( {t,\tau,{x_f}} \right) = {e_c}{\rm{exp}}\left( { - \frac{{{(t+\tau)^2}}}{{{T_f}^2}} - \frac{{{x_f}^2}}{{{w_f}^2}}} \right){\rm{exp}}\left[ {i\left( {{\omega _c} + \eta {x_f}} \right)(t+\tau)} \right]
	\end{equation}
Where $\eta  = 4\frac{{{w_i}\xi }}{{{w_f}{T_f}{T_i}}}$, $e_c (\omega _c)$ is the amplitude (frequency) of the control field, $w_i$ ($T_i$) is the spot size (duration) of the control field before focusing, $w_f$ ($T_f$) is the spot size (duration) of the control field at the focus, $\xi=\rho\xi_0$, $\rho$ is a constant, and $\xi_0$ is a spatiotemporal parameter. The perturbing phase can be written as:
	\begin{equation}	\label{equ7}
{\sigma ^j}\left( {{t_e},\tau ,{x_f}} \right) =  - \frac{9}{{128}}{e_c}{e_d}{\omega _d}t_e^4\exp \left( { - \frac{{{{\left( {\tau  } \right)}^2}}}{{T_f^2}} - \frac{{x_f^2}}{{w_f^2}}} \right){\rm{cos}}\left[ {\left( {{\omega _c} + \eta {x_f}} \right) {\tau} } \right]
	\end{equation}
When we use the control field in the noncollinear configuration:
	\begin{equation}	\label{equ8}
{{\rm{E}}_c}\left( {t,\tau,{x_f}} \right) = {e_c}{\rm{exp}}\left( { - \frac{{{{\left( {t+\tau + \frac{{{x_f}{\mathop{\rm sin}\nolimits} {\theta _c}}}{c}} \right)}^2}}}{{{T_f}^2}} - \frac{{{x_f}^2}}{{{w_f}^2}}} \right){\rm{exp}}\left[ {i{\omega _c}\left( {t+\tau + \frac{{{x_f}{\mathop{\rm sin}\nolimits} {\theta _c}}}{c}} \right)} \right]
	\end{equation}
Where $\theta_c$ is the angle between the control field and the driving field and $c$ is the velocity of light. Similarly, the perturbing phase is:
	\begin{equation}	\label{equ9}
\sigma^j \left( {{t_e},\tau ,{x_f}} \right) =  - \frac{9}{{128}}{e_c}{e_d}{\omega _d}t_e^4\exp \left( { - \frac{{{{\left( {\tau  + \frac{{{x_f}{\rm{sin}}{\theta _c}}}{c} } \right)}^2}}}{{T_f^2}} - \frac{{x_f^2}}{{w_f^2}}} \right){\rm{cos}}\left[ {{\omega _c}\left( {\tau  + \frac{{{x_f}{\rm{sin}}{\theta _c}}}{c}} \right)} \right]
	\end{equation}
When the high harmonic radiated from two slits are in a condition of constructive interference, the total phase difference is \cite{RN156}: 
	\begin{equation}\label{equ10}
	\Delta \phi \left( {\omega ,\tau ,{x_f}} \right) = {\sigma _2^j}\left( {{t_e},\tau ,{x_f}} \right) - {\sigma _1^j}\left( {{t_e},\tau ,{x_f}} \right) + \omega {T_0}/2 + \pi  + \gamma  = 2m\pi
	\end{equation}
Where $T_0$ is the optical cycle of the driving field, $\omega$ is the frequency of high harmonic, $\gamma$ is the non-adiabatic effect constant \cite{RN157}, and $m$ is an integer. Since the electric fields of neighboring time slits are reversed, their perturbing phases are also reversed: ${\sigma _1^j}\left( {{t_e},\tau ,{x_f}} \right) =  - {\sigma _2^j}\left( {{t_e},\tau ,{x_f}} \right)$. Then the perturbing phase of the first slit and second slit can be respectively written as:
	\begin{equation}\label{equ11}
	{\sigma_1^j} (t_e,\tau ,x_f) =  - \frac{9}{{128}}{e_d}\omega_d{t_e^4}E_c\left( {\tau ,x_f} \right)
	\end{equation}
	\begin{equation}\label{equ12}
	{\sigma_2^j} (t_e,\tau ,x_f) =  \frac{9}{{128}}\alpha(x_f){e_d}\omega_d{t_e^4}E_c\left( {\tau+\Delta(x_f) ,x_f} \right)
	\end{equation}
Where $\alpha(x_f)$ is the intensity ratio of the driving field between two neighboring slits, $\Delta(x_f)$ is the time difference between the two slits,
Combining Eqn.(\ref{equ10}), Eqn.(\ref{equ11}), and Eqn.(\ref{equ12}) :
	\begin{equation}\label{equ13}
	- 9/128{e_d}{\omega _d}t_e^4\left[ {\alpha(x_f) {E_c}\left( {\tau + \Delta(x_f) ,{x_f}} \right) + {E_c}\left( {\tau,{x_f}} \right)} \right] + \omega {T_0}/2 + \pi  + \gamma  = 2m\pi 
	\end{equation}
	\begin{equation}\label{equ14}
	\omega  = \left( {2m - 1} \right){\omega _d} - \frac{{2\gamma }}{{{T_0}}} + \frac{{9{e_d}{\omega _d}t_e^4}}{{64{T_0}}}\left[ {\alpha(x_f) {E_c}\left( {\tau + \Delta(x_f) ,{x_f}} \right) + {E_c}\left( {\tau ,{x_f}} \right)} \right]
	\end{equation}
For a certain order high harmonic, the emission time $t_e$ is certain. The spectral shift of a specific order high harmonic can be explained as:
	\begin{equation}\label{equ15}
	\delta\omega(\tau,{x_f})  = \frac{{9{e_d}{\omega _d}}}{{64{T_0}}}\left[ {\alpha(x_f) {E_c}\left( {\tau  + \Delta(x_f) ,{x_f}} \right) + {E_c}\left( {\tau  ,{x_f}} \right)} \right]
	\end{equation}	
When we introduce the control field with the WFR effect, we can substitute Eqn.(\ref{equ6}) into Eqn.(\ref{equ15}) and using the slow-varying envelope approximation: 
	\begin{equation}\label{equ16}
\delta \omega \left( {\tau ,{x_f}} \right) \propto \alpha ({x_f}){\rm{cos}}\left[ {\left( {{\omega _c} + \eta {x_f}} \right)\left( {\tau  + \Delta ({x_f})} \right)} \right] + {\rm{cos}}\left[ {\left( {{\omega _c} + \eta {x_f}} \right)\tau } \right]
	\end{equation}		
Notably, our simulation findings show that the delay and spatial dependent spectral shift of a specific harmonic order demonstrates a rather similar pattern for both near and far fields, see Fig.1(d) and (e) in the manuscript. Thus the far-field modulation of the HHs can be written as: 
	{\begin{equation}\label{equ17}
	\delta\omega(\tau,\theta) \propto {\alpha(\theta) {E_c}\left( {\tau  + \Delta(\theta) ,{\theta}} \right) + {E_c}\left( {\tau  ,{\theta}} \right)}.
	\end{equation}}
Because the coupling parameters for the first-order spatiotemporal coupling control field and noncollinear control fields are linearly distributed in space, the spatiotemporal coupling effect will be averaged out after the spatial integration. When we spatially integrate a high harmonic, the Eqn.(\ref{equ15}) turns to:
	\begin{equation}\label{equ18}
	\delta\omega(\tau)   \propto  {\alpha {E_c}\left( {\tau  + \Delta} \right) + {E_c}\left( {\tau} \right)} 
	\end{equation}
	
\par According to the Eqn.(\ref{equ5}), ${\sigma ^+} \propto  E_c\left( {\tau  ,\Delta{x_f}} \right)$, ${\sigma ^-} \propto E_c\left( {\tau  ,-\Delta{x_f}} \right)$. To the lowest order approximation, when we introduce the control field (using the slow-varying envelope approximation, $E_c ({\tau  ,\Delta{x_f}}) \propto \cos \left( {{\omega _c}\left( {\tau  + \frac{{\Delta {x_f}{\theta _c}}}{{{c}}}} \right)} \right)$) in the noncollinear configuration, the spatial shift in the far field can be explained as:
	\begin{equation}\label{equ19}
	\begin{aligned}
	\delta \theta  &\approx \frac{{{k_x}}}{{{k_z}}}\\
	&= \frac{1}{{{k_z}}}\frac{{\partial {\sigma ^{{j}}}}}{{\partial {{{x}}_f}}}\\
	&\approx \mathop {\lim }\limits_{\Delta {x_f} \to 0} \frac{1}{{{k_z}}}\frac{{{\sigma ^ + } - {\sigma ^ + }}}{{2\Delta {x_f}}}\\
	& \propto\mathop {\lim }\limits_{\Delta {x_f} \to 0}\frac{1}{{{k_z}}}\frac{{\cos \left( {{\omega _c}\left( {\tau  + \frac{{\Delta {x_f}{\theta _c}}}{{{c}}}} \right)} \right) - \cos \left( {{\omega _c}\left( {\tau  - \frac{{\Delta {x_f}{\theta _c}}}{{{c}}}} \right)} \right)}}{{2\Delta {x_f}}}\\
 	&= \mathop {\lim }\limits_{\Delta {x_f} \to 0}\frac{1}{{{k_z}}}\frac{{ - 2\sin \left( {\frac{{2{\omega _c}\tau }}{2}} \right)\sin \left( {\frac{{2{\omega _c}\Delta {x_f}{\theta _c}}}{{2c}}} \right)}}{{2\Delta {x_f}}}\\
 	&\approx \frac{{ - {\omega _c}{\theta _c}}}{{{k_z}c}}\sin \left( {{\omega _c}\tau } \right)\\
	& \propto {A_c}\left( \tau  \right)
	\end{aligned}
	\end{equation}
Similarly, when we introduce the control field with WFR, the spatial shift turns to:
	\begin{equation}\label{equ20}
	\begin{aligned}
	\delta \theta &\propto \mathop {\lim }\limits_{\Delta {x_f} \to 0}\frac{1}{{{k_z}}}\frac{{\cos \left( {\left( {{\omega _c} + \eta \Delta {{{x}}_f}} \right)\tau } \right) - \cos \left( {\left( {{\omega _c} - \eta \Delta {{{x}}_f}} \right)\tau } \right)}}{{2\Delta {x_f}}}\\
 	&= \mathop {\lim }\limits_{\Delta {x_f} \to 0}\frac{1}{{{k_z}}}\frac{{ - 2\sin \left( {\frac{{2{\omega _c}\tau }}{2}} \right)\sin \left( {\frac{{2\eta \Delta {{{x}}_f}\tau }}{2}} \right)}}{{2\Delta {x_f}}}\\
 	&\approx \frac{{ - \eta \tau }}{{{k_z}c}}\sin \left( {{\omega _c}\tau } \right)\\
	&\propto \eta \tau {A_c}\left( \tau  \right)
	\end{aligned}
	\end{equation}
However, when we fit the result of the simulation (see supplementary section 8), we find that Eqn.(\ref{equ20}) should shift a phase $\varphi_c$ and Eqn.(\ref{equ20}) turns to $\delta \theta \propto \eta \tau {A_c}(\tau + \varphi_c)$. Combine Eqn.(\ref{equ19}) with Eqn.(\ref{equ20}), we can explain the spatial shift when we introduce the control field with WFR in the noncollinear configuration:
	\begin{equation}\label{equ21}
	\begin{aligned}
	\delta \theta  &\propto  \mathop {\lim }\limits_{\Delta {x_f} \to 0}\frac{1}{{{k_z}}}\frac{{\cos \left( {\left( {{\omega _c} + \eta \Delta {{{x}}_f}} \right)\left( {\tau  + \frac{{\Delta {x_f}{\theta _c}}}{{{c}}}} \right)} \right) - \cos \left( {\left( {{\omega _c} - \eta \Delta {{{x}}_f}} \right)\left( {\tau  - \frac{{\Delta {x_f}{\theta _c}}}{{{c}}}} \right)} \right)}}{{2\Delta {x_f}}}\\
	& = \mathop {\lim }\limits_{\Delta {x_f} \to 0}\frac{1}{{{k_z}}}\frac{{ - 2\sin \left( {\frac{{2{\omega _c}\tau  + 2\eta \Delta x_f^2{\theta _c}/c}}{2}} \right)\sin \left( {\frac{{2\eta \Delta {{{x}}_f}\tau  + 2{\omega _c}\Delta {x_f}{\theta _c}/c}}{2}} \right)}}{{2\Delta {x_f}}}\\
	& \approx \frac{{ - \left( {\eta \tau  + {\omega _c}{\theta _c}/c} \right)}}{{{k_z}}}\sin \left( {{\omega _c}\tau } \right)\\
 	&\propto \kappa {A_c}\left( \tau  \right) + \eta \tau {A_c}\left( {\tau+ {\varphi _c}}  \right)
	\end{aligned}
	\end{equation}
Where $\kappa$ is related to the relative magnitude between the noncollinear angle and the value of WFR.

\section*{Section 2: The time window function and the simulation of high harmonic spectrum}

\begin{figure}[H]
	\centering
	\includegraphics[width=.5\linewidth]{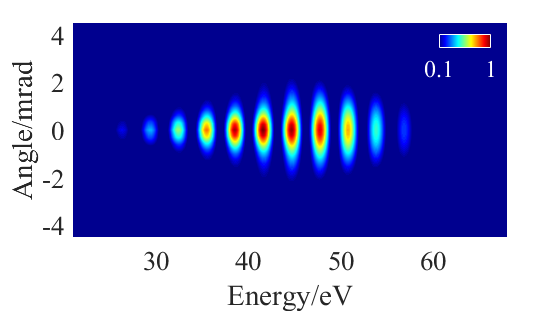}
	\caption{The simulation of high harmonic spectrum in the far field.}
	\label{FigureS1}
	\end{figure}

\par Here we add a series of time window functions (super-Gaussian window $\exp \left[ { - {{\left( {\frac{{{\chi _1}\Delta t}}{{{\chi _2}T_0}}} \right)}^4}} \right]$, $\chi _1$ and $\chi _2$ is adjustment parameters of the functions, $\Delta t$ is the width of the windows) on the radiation dipole moment of electron for selecting the high harmonics in short trajectories region and Fig.{\ref{FigureS1}} is the high harmonic spectrum after selection in the far field. 

\section*{Section 3: Experimental setup}
	
	\begin{figure}[H]
	\centering
	\includegraphics[width=1\linewidth]{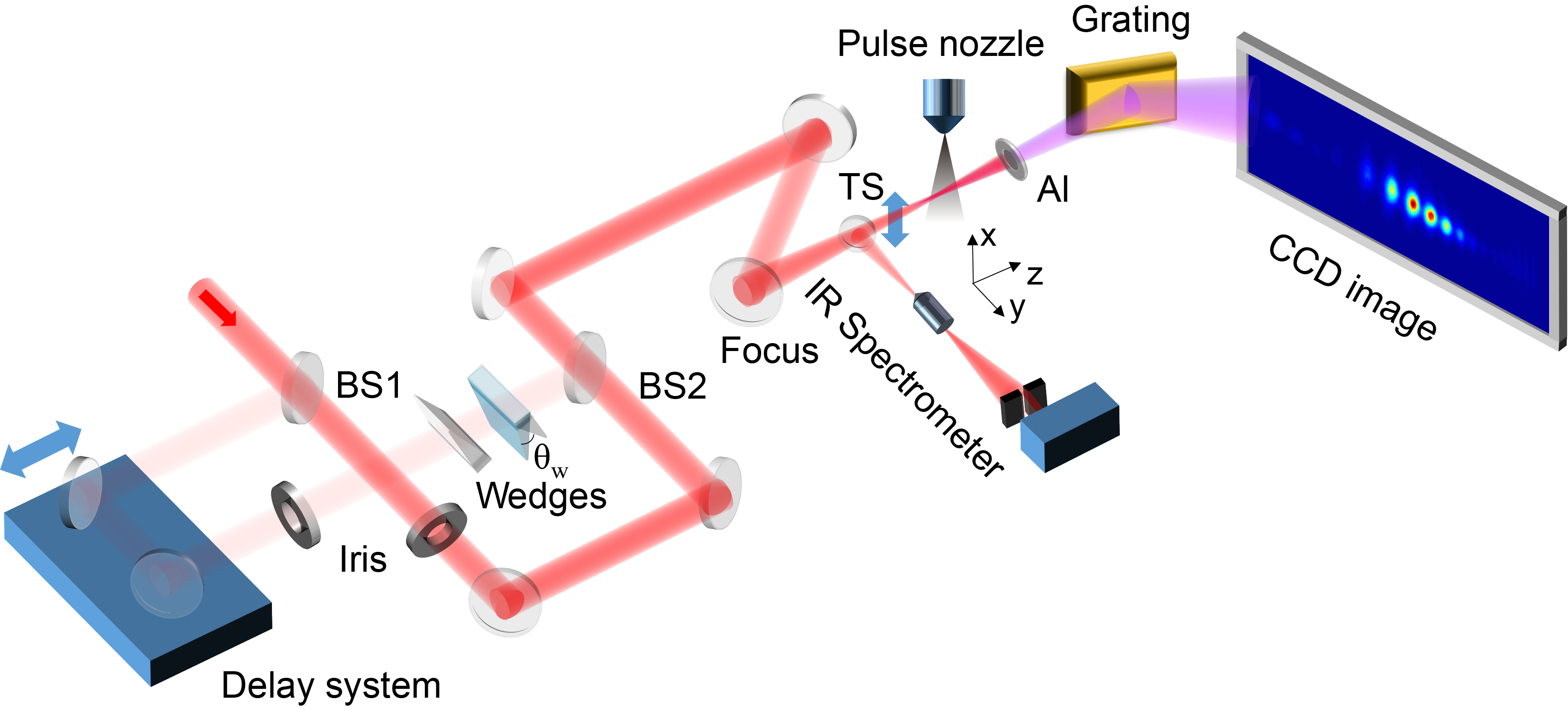}
	\caption{Experimental setup diagram. BS is a beam splitter, TS is a thruster, and Al is aluminum film.}
	\label{FigureS2}
	\end{figure}

\par Our device is shown in Fig.{\ref{FigureS2}}. We use fiber filled with 1.2 bar neon gas and chirped mirrors to compress the pulse to around 11 fs. The pulse laser has a central wavelength of 780 nm, and a repetition frequency of 1000 hz. The pulse is split by a 20$\%$ reflection beam splitter. The transmission beam serves as a driving field. A delay stage and iris are inserted in the reflection beam path to adjust the delay between two pulses and the intensity of the control field, respectively. A pair of misaligned wedges (top angle is $8^{\circ}$,  misaligned angle is $\theta_w$) is used to introduce the pulse front tilt effect to the control field and after focusing the control field will carry the spatial chirp at the focus. After that, two beams are recombined by another 20$\%$ reflection beam splitter and focused to interact with argon gas injected from a pulsed nozzle with an aperture of 220 ${\mu}$m and a back-pressure of 20 bar to generate high harmonics. After passing through the aluminum film which is used to filter the NIR pulses and the grating, the high harmonics are detected on the CCD. An additional reflector fixed to the thruster can reflect the beam to the IR spectrometer which is combined with a 10$\times$ microscope objective, a slit, and a spectrometer and we can use it to measure the spatial spectrum of the control pulse. 
	
	\section*{Section 4: The spatial spectrum of the control field}
	
	\begin{figure}[H]
	\centering
	\includegraphics[width=.5\linewidth]{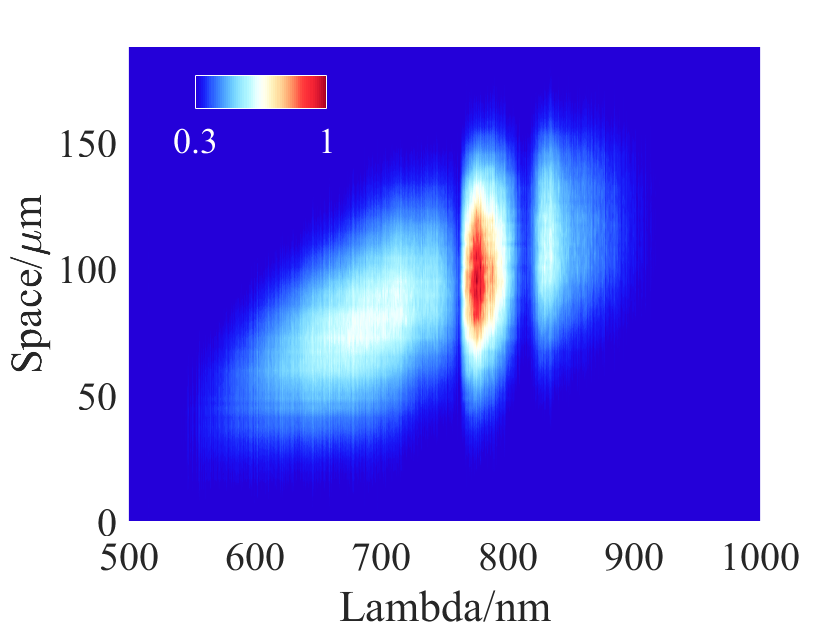}
	\caption{The spatial spectrum of the control pulse at the focus.}
	\label{FigureS3}
	\end{figure}

\par We can measure the x-direction spatial spectrum of the control pulse by the IR spectrometer and the result is shown as Fig.{\ref{FigureS3}}. It shows that the control field contains the spatial chirp at the focus, which is consistent with others’ results \cite{RN150}.
	
	\section*{Section 5: Comparison of the simulation result of the spectral shift}

	\begin{figure}[H]
	\centering
	\includegraphics[width=.5\linewidth]{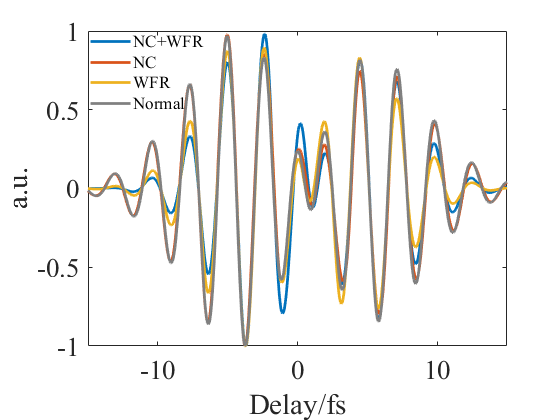}
	\caption{The simulation results of the spectral shift of $33^{rd}$ order high harmonic for different conditions of the control field. NC: noncollinear configuration; Normal: collinear plane wave.}
	\label{FigureS4}
	\end{figure}

The spectral shifts of four different conditions of the control field are almost the same except that the amplitude is slightly modulated.

\section*{Section 6: The fitting result of the ellipse}

	\begin{figure}[H]
	\centering
	\includegraphics[width=.5\linewidth]{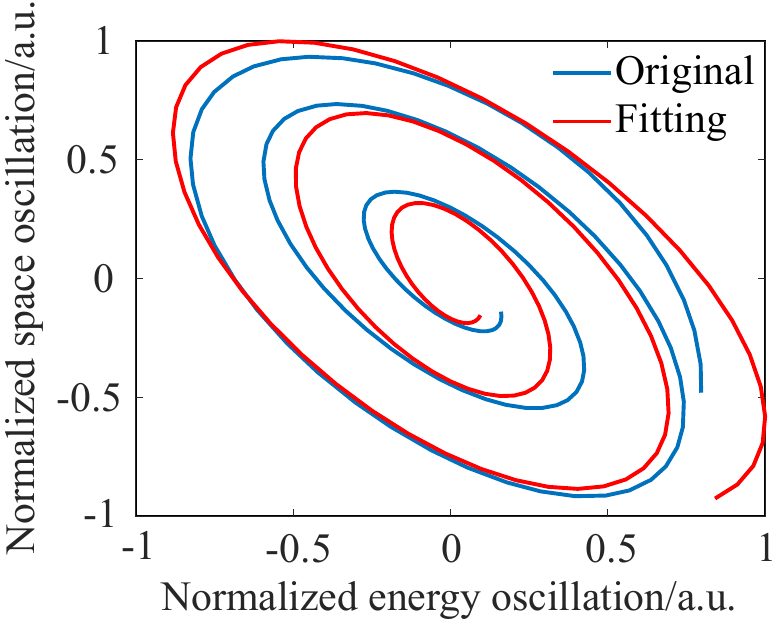}
	\caption{The fitting result of the ellipse. The blue line is the numerical simulation result of the $33^{rd}$ order high harmonic for the control field with WFR and the red line is the fitting result.}
	\label{FigureS5}
	\end{figure}
	
\par After using a trigonometric curve fitting of modulations, we obtain the value of the constant phase, $\varphi_c=-0.196\pi$, and we can see they match well.

	\section*{Section 7: Other results of the simulation}

	\begin{figure}[H]
	\centering
	\includegraphics[width=1\linewidth]{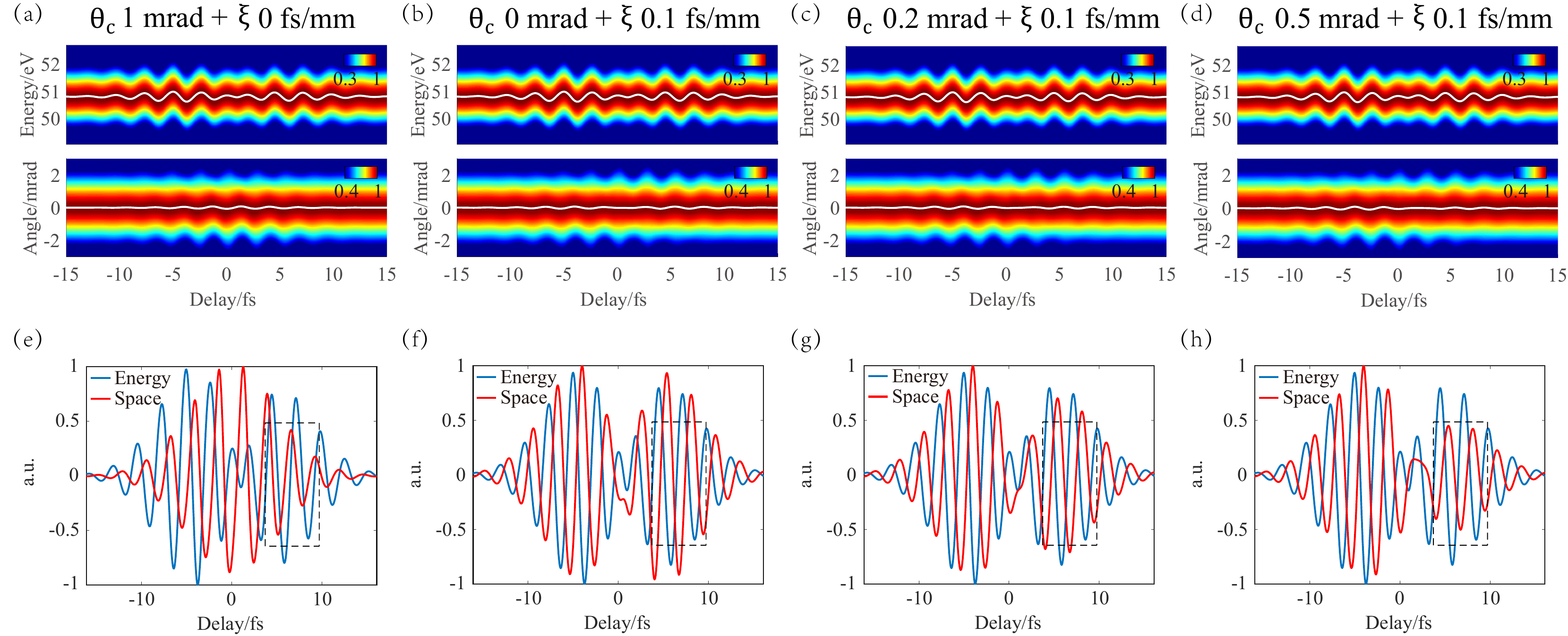}
	\caption{(a)-(d) The simulated spatially integrated spectrum (upper row) and spectrally integrated spatial distribution (lower row) of the $33^{rd}$ order harmonic under different control field conditions. The white solid lines show the center of mass of the modulation. (e)-(h) The normalized center of mass curves from (a)-(d).}
	\label{FigureS6}
	\end{figure}

\section*{Section 8: Modulation of high harmonic with second-order spatiotemporal coupling effect}

	\begin{figure}[H]
	\centering
	\includegraphics[width=0.5\linewidth]{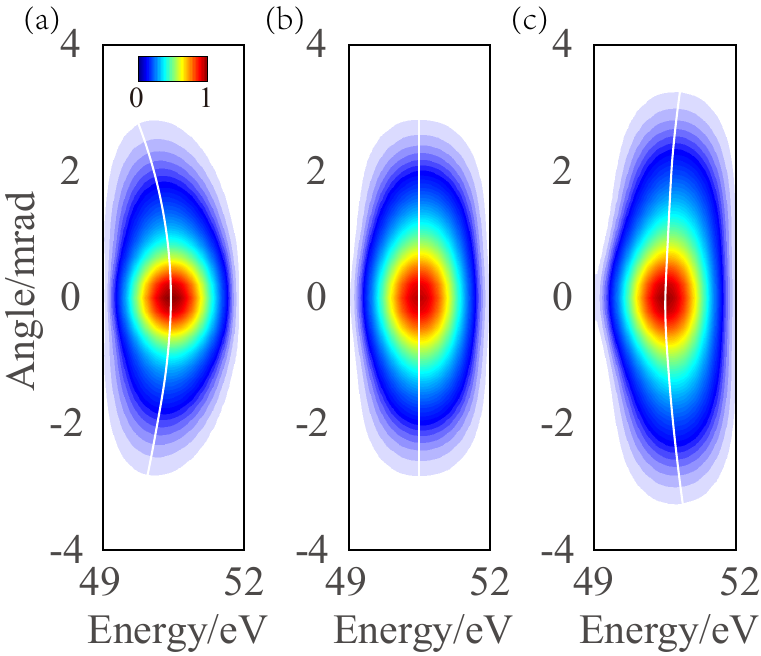}
	\caption{The spatio-spectral beam profiles of the $33^{rd}$ order harmonic for delays: (a) -1.3 fs, (b) -0.7 fs, and (c) 0 fs. The control field has a second order spatiotemporal coupling laser field with the form: ${{\rm{E}}_c}\left( {t ,{x_f}} \right) = {e_c}\cos \left[ {\left( {{\omega _c} + \eta {x_f}^2} \right)t } \right]$}
	\label{FigureS7}
	\end{figure}

	\section*{Section 9: The simulation result of different orders of high harmonics}

	\begin{figure}[H]
	\centering
	\includegraphics[width=.5\linewidth]{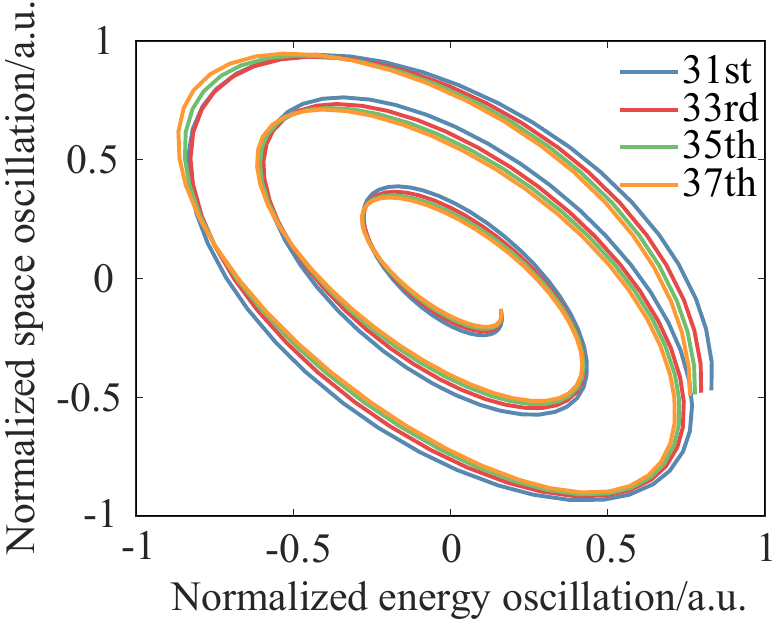}
	\caption{The simulated ($\delta \omega$,$\delta \theta$) ellipses for different orders of harmonic using the same WFR control field. $\delta \omega$: the spectral shift; $\delta \theta$: the emission angle.}
	\label{FigureS8}
	\end{figure}

\par Different orders of high harmonics have the same behavior under the same condition of the control field.

	\section*{Section 10: The simulation result for the control field in the collinear structure and with different values of WFR}

	\begin{figure}[H]
	\centering
	\includegraphics[width=.5\linewidth]{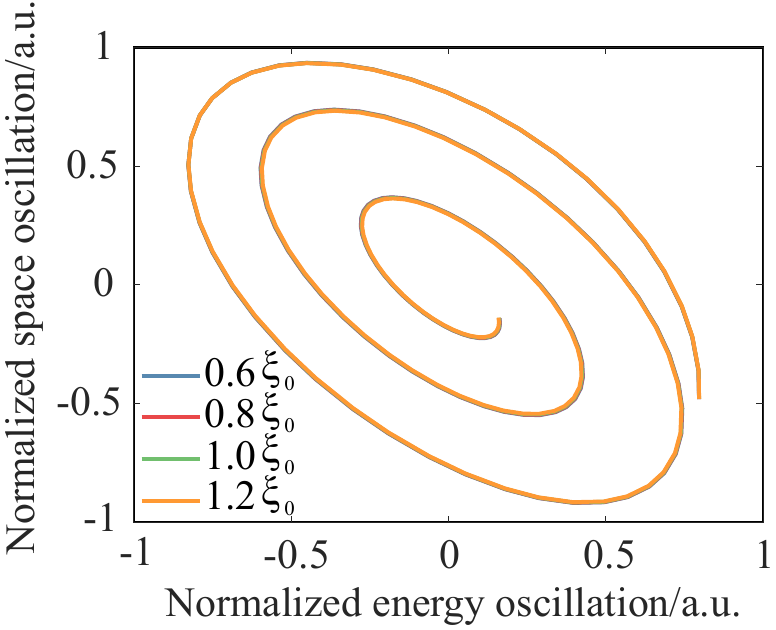}
	\caption{Same as Fig.{\ref{FigureS8}}, but under a collinear configuration with different values of WFR in the control field.}
	\label{FigureS9}
	\end{figure}

\par When we change the values of WFR of the control field in the collinear configuration, the high harmonic has the same behavior.

	\section*{Section 11: The simulation result for the control field with the same noncollinear angle but different values of WFR}

	\begin{figure}[H]
	\centering
	\includegraphics[width=.5\linewidth]{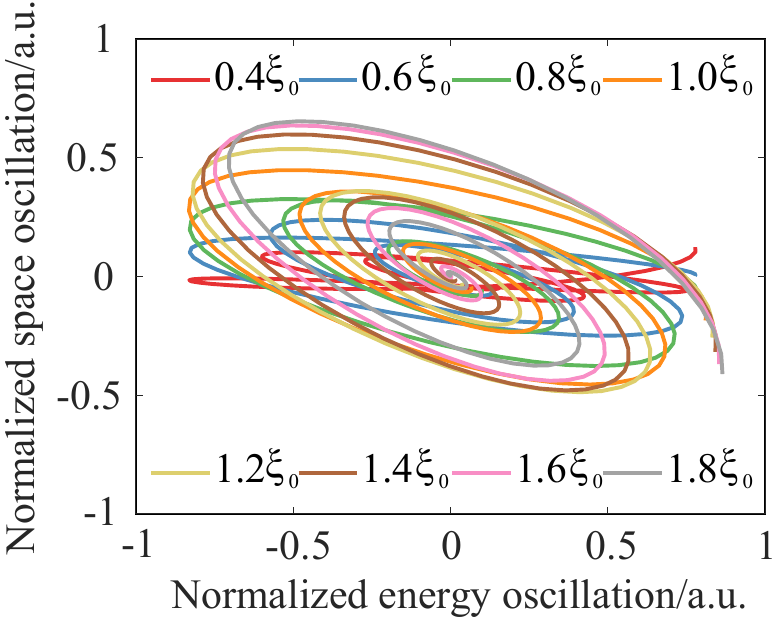}
	\caption{Same as Fig.{\ref{FigureS8}}, but under a noncollinear configuration with different values of WFR in the control field.}
	\label{FigureS10}
	\end{figure}

\par When we change the values of WFR of the control field with the same noncollinear angle, the tilt angle of the ellipse changes. When the noncollinear angle is fixed, the larger the value of WFR is, the larger the tilt angle of the ellipse is.